\documentclass[12pt,reqno]{amsart}
\usepackage{graphicx}
\usepackage{amscd}
\usepackage{amsmath}
\usepackage{epsfig}
\usepackage{amsfonts}
\usepackage{amssymb}

\providecommand{\U}[1]{\protect\rule{.1in}{.1in}}
\providecommand{\U}[1]{\protect\rule{.1in}{.1in}}
\allowdisplaybreaks
\newtheorem{theorem}{Theorem}
\theoremstyle{plain}

\newtheorem{corollary}{Corollary}

\newtheorem{definition}{Definition}

\newtheorem{lemma}{Lemma}

\newtheorem{proposition}{Proposition}
\newtheorem{remark}{Remark}

\numberwithin{equation}{section}

\begin{document}
\title[Schr\"{o}dinger Equation]{On Schr\"{o}dinger Equation with
Time-Dependent Quadratic Hamiltonian in $R^d$}
\author{Erwin Suazo}
\address{Department of Mathematics and Statistics, Arizona State University,
Tempe, AZ 85287--1804, U.S.A.}
\email{esuazo@asu.edu}
\urladdr{http://mathpost.asu.edu/\symbol{126}suazo/index.html}

\begin{abstract}
We study solutions to the Cauchy problem for the linear and nonlinear Schr%
\"{o}dinger equation with a quadratic Hamiltonian depending on time. For the
linear case the evolution operator can be expressed as an integral operator
with the explicit formula for the kernel. As a consequence, conditions for
local and global in time Strichartz estimates can be established. For the
nonlinear case we show local well-posedness. As a particular case we obtain
well-posedness for the damped harmonic nonlinear Schr\"{o}dinger equation.
\end{abstract}

\maketitle



\section{\textbf{INTRODUCTION}}

In this paper we first study the time-dependent linear Schr\"{o}dinger
equation
\begin{equation}
i\frac{\partial \psi }{\partial t}=H\left( t\right) \psi ,\qquad \psi
(x,0)=\varphi  \label{LStime}
\end{equation}%
with the quadratic Hamiltonian%
\begin{equation}
H(t)\psi =-\frac{1}{2}\Delta \psi +\sum_{j=1}^{d}\left( \frac{b_{j}\left(
t\right) }{2}x_{j}^{2}\psi -f_{j}(t)x_{j}\psi +ig_{j}(t)\frac{\partial \psi
}{\partial x_{j}}-i\frac{c_{j}\left( t\right) }{2}\left( 2x_{j}\frac{%
\partial \psi }{\partial x_{j}}+\psi \right) \right) ,  \label{Hamiltonian}
\end{equation}%
where $b_{j},f_{j},g_{j},c_{j}\in C^{1}$ ($b_{j},f_{j},g_{j}$ could be
piecewise continuous functions) and $\varphi \in S(\mathbb{R}^{d})$ ($S(%
\mathbb{R}^{d})$ is the Schwartz space) to simplify the discussion. We
derive its evolution operator given by an explicit formula in the form
\begin{equation}
\psi \left( x,t\right) =U_{H}(t,0)\varphi \left( x\right) =\int_{\mathbb{R}%
^{d}}G_{H}(x,y,t)\ \varphi \left( y\right) \ dy  \label{integral operator}
\end{equation}%
by constructing the fundamental solution (FS) associated to (\ref{LStime})-(%
\ref{Hamiltonian}), see Lemma 1; the FS is a solution of (\ref{LStime}) with
the initial data $G_{H}(x,y,0)=\delta (x-y)$. Next we study properties of $%
U_{H}$ and give conditions to obtain local and global in time Strichartz
estimates for $\varphi \in L_{x}^{2}(\mathbb{R}^{d})$. We prove that the
nonlinear version of (\ref{LStime})-(\ref{Hamiltonian}) with algebraic
nonlinearity is locally wellposed in $L_{x}^{2}(\mathbb{R}^{d})$ in the
subcritical sense, see Section 4. Finally, we introduce the damped harmonic
nonlinear Schr\"{o}dinger equation:
\begin{equation}
i\frac{\partial u}{\partial t}=\frac{\omega _{0}}{2}\left( -\frac{\partial
^{2}u}{\partial x^{2}}+x^{2}u\right) +i\frac{\lambda }{2}\left( 2x\frac{%
\partial u}{\partial x}+u\right) +h|u|^{p-1}u
\label{damped harmonic nonlinear Schr}
\end{equation}%
and also prove well-posedness. A systematic study of the blow up and
scattering results of (\ref{LStime})-(\ref{Hamiltonian}) are presented in
\cite{Sua:Sus2}.

The study of time dependent quadratic Hamiltonians is quite complex and has
brought a great deal of attention see \cite{Hagedorn1998}, \cite%
{UmYeonGeorge} and references therein, see also \cite{Hagedorn1986} for a
nice discussion on this type of Hamiltonians consider \textquotedblleft folk
wisdow". The study of methods (i.e. propagator method, dynamical invariant
and second quatization methods) used to find the exact propagators for Schr%
\"{o}dinger equations has also been studied by several authors \cite%
{UmYeonGeorge}. The equation (\ref{LStime}) can be solved, at least
formally, using a time evolution operator, $U\left( t,t_{0}\right) $ given by%
\begin{equation}
U\left( t,t_{0}\right)
=\sum_{k=0}^{n}(-i)^{k}\int_{t_{0}}^{t_{1}}dt_{1}...%
\int_{t_{0}}^{t_{k-1}}dt_{k}H(t_{1})...H(t_{k})=\text{T}\left( \exp \left( -%
\frac{i}{\hslash }\int_{t_{0}}^{t}H\left( t^{\prime }\right) \ dt^{\prime
}\right) \right) ,
\end{equation}%
where T is the time ordering operator which orders operators with larger
times to the left, and of course this expression might diverge.

The fundamental solution $G_{H}$ for the equation (\ref{LStime})-(\ref%
{Hamiltonian}) includes the following examples with the explicit expressions:

Table I. Some exactly solvable quadratic Hamiltonians (We assume $E,$ $k$
constants).

\begin{tabular}{|l|l|}
\hline
Hamiltonian $H(t)$ & Fundamental Solution (Propagator) \\ \hline
$%
\begin{array}{l}
\text{Free Particle} \\
H_{0}(t)\psi =-\frac{1}{2}\frac{\partial ^{2}\psi }{\partial x^{2}}%
\end{array}%
$ & $G_{0}(x,y,t)=\frac{1}{\sqrt{2\pi i\sin t}}e^{i\mid x-y\mid ^{2}/2t}$ \\
\hline
$%
\begin{array}{l}
\text{ Constant Electric Field } \\
\text{ }H_{1}(t)\psi =-\frac{1}{2}\frac{\partial ^{2}\psi }{\partial x^{2}}%
+E\cdot x\psi%
\end{array}%
$ & $%
\begin{array}{l}
G_{1}\left( x,y,t\right) =\frac{1}{\sqrt{2\pi i\sin t}}\exp \left( \frac{%
i\left( x-y\right) ^{2}}{2t}\right) \\
\times \exp \left( \frac{iE\left( x+y\right) }{2}t-\frac{iE^{2}}{24}%
t^{3}\right)%
\end{array}%
$ \\ \hline
$%
\begin{array}{l}
\text{ Isotropic Oscillator} \\
\text{ }H_{2}(t)\psi =-\frac{1}{2}\frac{\partial ^{2}\psi }{\partial x^{2}}+%
\frac{1}{2}x^{2}\psi%
\end{array}%
$ & $%
\begin{array}{l}
G_{2}(x,y,t)=\frac{1}{\sqrt{2\pi i\sin t}} \\
\times \exp \left( i\frac{1}{4{\sin (t)}}\left( \left( x^{2}+y^{2}\right)
\cos t-2xy\right) \right)%
\end{array}%
$ \\ \hline
$%
\begin{array}{l}
\text{Repulsive harmonic potential} \\
\text{ }H_{3}(t)\psi =-\frac{1}{2}\frac{\partial ^{2}\psi }{\partial x^{2}}-%
\frac{1}{2}x^{2}\psi%
\end{array}%
$ & $%
\begin{array}{l}
G_{3}(x,y,t)=\frac{1}{\sqrt{2\pi i\sinh t}} \\
\times \exp \left( i\frac{1}{4{\sinh t}}\left( \left( x^{2}+y^{2}\right)
\cosh t-2xy\right) \right)%
\end{array}%
$ \\ \hline
$%
\begin{array}{l}
\text{ Anisotropic Oscillator} \\
\text{ }H_{4}(t)\psi =-\frac{1}{2}\frac{\partial ^{2}\psi }{\partial x^{2}}+%
\frac{1}{2}\omega ^{2}x^{2}\psi%
\end{array}%
$ & $%
\begin{array}{l}
G_{4}(x,y,t)=\frac{\omega }{\sqrt{2\pi i\sin \omega t}} \\
\times \exp \left( i\frac{\omega }{4{\sin (\omega t)}}\left( \left(
x^{2}+y^{2}\right) \cos \omega t-2xy\right) \right)%
\end{array}%
$ \\ \hline
$%
\begin{array}{l}
\text{ Modified MC-SS Oscillator } \\
\text{ }H_{6}(t)\psi =-\cos ^{2}t\frac{\partial ^{2}\psi }{\partial x^{2}}%
+\sin ^{2}tx^{2}\psi \\
\qquad \qquad -i\frac{\sin 2t}{2}\left( 2x\frac{\partial }{\partial x}%
-1\right) \psi%
\end{array}%
$ & $%
\begin{array}{l}
G_{6}\left( x,y,t\right) =\frac{1}{\sqrt{2\pi i\left( \cos t\sinh t+\sin
t\cosh t\right) }} \\
\times \exp \left( \frac{\left( x^{2}-y^{2}\right) \sin t\sinh t+2xy-\left(
x^{2}+y^{2}\right) \cos t\cosh t}{2i\left( \cos t\sinh t+\sin t\cosh
t\right) }\right)%
\end{array}%
$ \\ \hline
$%
\begin{array}{l}
\text{Damped Harmonic Oscillator } \\
\text{ }H_{7}(t)\psi =-\frac{\omega _{0}}{2}\frac{\partial ^{2}\psi }{%
\partial x^{2}}+\frac{\omega _{0}}{2}x^{2}\psi \\
+i\frac{\lambda }{2}\left( 2x\frac{\partial }{\partial x}+1\right) \psi%
\end{array}%
$ & $%
\begin{array}{l}
G_{7}\left( x,y,t\right) =\sqrt{\frac{\omega }{2\pi i\omega _{0}\sin \omega t%
}} \\
\times \exp \left( \frac{i\omega }{2\omega _{0}\sin \omega t}\left( \left(
x^{2}+y^{2}\right) \cos \omega t-2xy\right) \right) \\
\times \exp \left( \frac{i\lambda }{2\omega _{0}}\left( x^{2}-y^{2}\right)
\right) ,\text{ }\omega =\sqrt{\omega _{0}^{2}-\lambda ^{2}}>0%
\end{array}%
$ \\ \hline
$%
\begin{array}{l}
\text{Analog of Heat Equation} \\
\text{ with Linear Drift} \\
\text{ }H_{8}(t)\psi =-\frac{\partial ^{2}\psi }{\partial x^{2}}-ikx\frac{%
\partial \psi }{\partial x},\qquad k>0%
\end{array}%
$ & $G_{8}(x,y,t)=\frac{\sqrt{k}e^{kt/2}}{\sqrt{2\pi i\sinh (kt)}}\exp
\left( \frac{ike^{kt}\left[ e^{-kt}x-e^{kt}y\right] ^{2}}{4\sinh (kt)}%
\right) $ \\ \hline
\end{tabular}

The\ generality of equation (\ref{LStime})-(\ref{Hamiltonian}) includes
several examples of interest not only for linear \cite{Fey:Hib}, \cite%
{UmYeonGeorge}, but for nonlinear Schr\"{o}dinger equations (NLS) \cite%
{Carles}, \cite{Carlesstark}; for a general review in NLS see \cite{Caz},
\cite{Tao}. The explicit solution formula for (\ref{LStime})-(\ref%
{Hamiltonian}), see (\ref{General Operator}), allows us to treat nonlinear
versions of these cases with the advantage of allowing time dependence
factors so that questions such as local and global existence, finite time
blow up and scattering can be solved using similar methods to those used in
the case of the nonlinear Schr\"{o}dinger equation without potential. The
results presented here are of interest to study the behavior in time of
singularities of solutions for Schr\"{o}dinger equations in the same
direction as in \cite{Zelditch}. Also the fact that in quantum mechanics it
is rare to find an exact solution to nonstationary problems, see \cite%
{Peremolov:Zel}, \cite{Peremolov:Popov}, makes our explicit solution useful
for testing numerical methods to solve time-dependent Schr\"{o}dinger
equation. One of the original motivations of this paper is to introduce a
generalization of the practical formula (\ref{formula1})-(\ref{formula2}),
for applications see \cite{Carles}, \cite{Zelditch} and \cite{Visan}.

\subsection{Quadratic Quantum Hamiltonians}

The expert would recognize (\ref{Hamiltonian}) as a quantum mechanical
self-adjoint Hamiltonian, which is a quadratic polynomial in $x$ and $%
p=-i\partial /\partial x$ with time dependent coefficients \cite%
{Hagedorn1986}, \cite{Hagedorn1998}:
\begin{equation}
H(t)=a(t)p^{2}+\frac{c\left( t\right) }{2}\left( p\cdot x+x\cdot p\right) +%
\frac{b\left( t\right) }{2}x^{2}-g(t)p-f(t)x+\zeta (t).
\end{equation}%
As pointed out in \cite{Hagedorn1986} one can assume $\zeta (t)=0$ since it
causes a trivial phase factor in the propagator. We also have assumed in (%
\ref{LStime}) $a(t)=1/2,$ thinking we can disregard it after a substitution,
however there are important cases that we will be missing, for example the
Caldirola-Kanai Hamiltonian \cite{Caldirola}, \cite{Kanai}. In \cite%
{Co:Suazo:Su} the authors did not know about this case and called it
\textquotedblleft third model," it was one the models of the damped harmonic
oscillator with explicit propagator considered in that publication. The
Caldirola-Kanai Hamiltonian was introduced more than 60 years ago \cite%
{Caldirola}, \cite{Kanai}:
\begin{equation}
\text{ }H_{CK}(t)\psi =-\frac{\omega _{0}e^{-\lambda t}}{2}\frac{\partial
^{2}\psi }{\partial x^{2}}+\frac{\omega _{0}e^{\lambda t}}{2}x^{2}\psi .
\tag{Caldirola-Kanai Hamiltonian}
\end{equation}%
its fundamental solution is given by%
\begin{equation}
G_{CK}(x,y,t)=\sqrt{\frac{\omega e^{\lambda t}}{2\pi i\omega _{0}\sin \omega
t}}e^{i(\alpha (t)x^{2}+\beta (t)xy+\gamma (t)y^{2})},
\end{equation}%
where

\begin{eqnarray}
\alpha (t) &=&\frac{\omega \cos \omega t-\lambda \sin \omega t}{2\omega
_{0}\sin \omega t}e^{2\lambda t}, \\
\text{ }\beta (t) &=&-\frac{\omega }{\omega _{0}\sin \omega t}e^{\lambda t},%
\text{ } \\
\gamma (t) &=&\frac{\omega \cos \omega t+\lambda \sin \omega t}{2\omega
_{0}\sin \omega t}.
\end{eqnarray}%
This model has been studied for several authors, and a detailed bibliography
can be found in \cite{UmYeonGeorge}.

Another example with $a(t)$ not constant that can be solved using the same
ideas presented in this work is%
\begin{equation}
H_{10}(t)\psi =-a(t)\frac{\partial ^{2}\psi }{\partial x^{2}}+\frac{b(t)}{2}%
x^{2}\psi ,
\end{equation}%
\begin{eqnarray}
a(t) &=&\frac{\left( \Omega ^{2}\cos (\Omega t)-\gamma \sin (\Omega t)\tanh
(\gamma t)\right) }{\cosh (\gamma t)(\cos (\gamma t)\cosh (\gamma t)-2\gamma
)},\text{ } \\
b(t) &=&-\frac{\omega ^{2}}{4a(t)},\text{ }\Omega =\sqrt{\omega ^{2}-\gamma
^{2}},
\end{eqnarray}%
and its fundamental solution is given by%
\begin{equation}
G_{10}(x,y,t)=\sqrt{\frac{m_{0}\Omega \cosh \gamma t}{2\pi i\sin (\Omega t)}}%
e^{i(\alpha (t)x^{2}+\beta (t)xy+\gamma (t)y^{2})},
\end{equation}%
with

\begin{equation}
\alpha (t)=\frac{\cosh (\gamma t)\left( m_{0}\Omega \cosh (\gamma t)\cos
(\Omega t)-\gamma \right) }{2\sin (\Omega t)},
\end{equation}%
\begin{equation}
\beta (t)=-\frac{m_{0}\Omega \cosh \gamma t}{2\pi i\sin (\Omega t)},
\end{equation}%
\begin{equation}
\gamma (t)=\frac{m_{0}\Omega \cos (\Omega t)}{2\sin (\Omega t)}.
\end{equation}%
Suslov et. al in \cite{Co:Suazo:Su2} \ study the quantum integrals of motion
between others of these last two models.

\subsection{A General Formula}

If we consider the linear Schr\"{o}dinger equation on $\mathbb{R}^{n}$ (\ref%
{LStime})-(\ref{Hamiltonian}) with%
\begin{equation}
H_{V}(t)\psi =-\frac{1}{2}\Delta \psi +V(x)\psi ,  \label{linear carles}
\end{equation}%
where $V(x)=\sum_{j=1}^{n}\left( \delta _{j}\frac{\omega _{j}^{2}}{2}%
x_{j}^{2}+b_{j}x_{j}\right) $, $n\geq 1,$ $\omega _{j}>0,$ $\delta _{j}\in $
$\{-1,0,1\},$ then the solution is given by the following formula \cite%
{Carles}
\begin{equation}
\psi (x,t)=U_{V}(t)f:=\frac{1}{\sqrt{2i\pi g_{j}(t)}}\int_{\mathbb{R}%
^{n}}e^{iS_{V}(x,y,t)}f(y)dy  \label{formula1}
\end{equation}%
where
\begin{equation*}
S_{V}(x,y,t)=\frac{1}{g_{j}(t)}\left( \frac{x_{j}^{2}+y_{j}^{2}}{2}%
h_{j}(t)-x_{j}y_{j}\right)
\end{equation*}%
\begin{equation}
\{g_{j}(t),h_{j}(t)\}=\left\{
\begin{tabular}{ll}
$\left\{ \frac{\sinh (\omega _{j}t)}{\omega _{j}},\cosh (\omega
_{j}t)\right\} $, & $\text{if }\delta _{j}=-1,$ \\
$\ \ \ \ \ \ \ \ \ \ \ \ \ \ \ \ \ \ \ \ \{t,1\},$ & $\text{if }\delta
_{j}=0,$ \\
$\ \ \left\{ \frac{\sin (\omega _{j}t)}{\omega _{j}},\cos (\omega
_{j}t)\right\} ,$ & $\text{if }\delta _{j}=+1.$%
\end{tabular}%
\right.  \label{formula2}
\end{equation}

This formula includes the free particle propagator introduced by Ehrenfest
\cite{Ehrenfest} that corresponds to the case $\delta _{j}=0$ for all $j$ in
(\ref{linear carles}). The propagator for the simple harmonic oscillator is
obtained from (\ref{formula1}), by choosing $\delta _{j}=1$ for all $j$ in (%
\ref{linear carles}) (a consequence of Mehler's formula for Hermite
polynomials \cite{Fey:Hib}). Finally, the propagator for the isotropic
harmonic potential is obtained by choosing $\delta _{j}=-1$ for all $j$ in (%
\ref{linear carles}).

The fact that we want to emphasize in formulas (\ref{formula1})-(\ref%
{formula2}) is that it combines the fundamental solutions of the free
particle and simple and isotropic harmonic oscillators in the one
dimensional case (using tensor product) to construct the explicit solutions
from a variety of Schr\"{o}dinger equations in\ several dimensions. One
motivation comes from the Schr\"{o}dinger equation (\ref{LStime})-(\ref%
{Hamiltonian}) with the Hamiltonian%
\begin{equation}
H_{h}(t)\psi =-\frac{1}{2}\Delta \psi +1/2\left( -\omega
_{1}x_{1}^{2}+\omega _{2}x_{2}^{2}\right) \psi ,
\end{equation}%
(compare with $H_{3}$ and $H_{4}$ from Table I) whose evolution operator
satisfies global in time Strichartz estimates, see \cite{Carles}.

However, the explicit formula for the evolution operator corresponding to\ (%
\ref{LStime})-(\ref{Hamiltonian}) allows us to consider a wider range of
operators to solve explicitly a variety of Schr\"{o}dinger equations in
several dimensions.

\subsection{Cases from Table I Not Included in Formula (\protect\ref%
{formula1})-(\protect\ref{formula2})}

The case of the particle in a constant external field having the Hamiltonian
$H_{1}(t)$ case was studied in detail by \cite{Cycon:Froese:Kirsch:Simon},
\cite{Arrighini:Durante}, \cite{Brown:Zhang}, \cite{Fey:Hib}, \cite%
{Holstein97}, \cite{Nardone} and \cite{Robinett}. The forced harmonic
oscillator is obtained in (\ref{LStime})-(\ref{Hamiltonian}) by choosing
\begin{equation*}
H(t)\psi =\frac{\hbar \omega }{2}\left( -\frac{\partial ^{2}}{\partial x^{2}}%
+x^{2}\right) \psi +\frac{\hbar }{\sqrt{2}}\left( \delta (t)\left( x-\frac{%
\partial }{\partial x}\right) +\delta ^{\ast }(t)\left( x+\frac{\partial }{%
\partial x}\right) \right) \psi ,
\end{equation*}%
where $\delta (t)$ is a complex valued function of time $t$ and the symbol $%
\ast $ denotes complex conjugation. It corresponds to the case which is of
interest in many advanced problems, examples including polyatomic molecules
in varying external fields, crystals through which an electron is passing
and exciting the oscillator modes, and other interactions of the modes with
external fields. An example of this type of Hamiltonian is $H_{6}(t),$
introduced in \cite{Me:Co:Su}.

The damped oscillations have been analyzed to a great extent in classical
mechanics; see, for example, \cite{BatemanPDE} and \cite{Lan:Lif}. In \cite%
{Co:Suazo:Su} the damped harmonic Schr\"{o}dinger equation with self-adjoint
Hamiltonian $H_{7}(t)$ was considered. The case $H_{8}(t)$ is taken from the
analogous case of the heat equation with linear drift, see \cite{Miller} for
a proof of this fundamental solution using Lie theory approach.

All examples in Table I are explicit solutions because one can solve the
characteristic equation (\ref{characteristic}) and evaluate the integrals (%
\ref{schr17})-(\ref{schr19}) explicitly. Solving (\ref{characteristic})
could require elaborated techniques, see for example \cite{Nathan:Suslov}.
However, even without solving (\ref{characteristic}) explicitly important
conclusions can be drawn. For example, see \cite{Carles3}, where%
\begin{equation}
H_{\tilde{V}}(t)\psi =-\frac{1}{2}\Delta \psi +\tilde{V}(x,t)\psi
\end{equation}%
with $\tilde{V}(x,t)\psi =\sum_{j=1}^{d}\frac{b_{j}\left( t\right) }{2}%
x_{j}^{2}\psi $ was considered and finite time blow up results are obtained$%
. $ Another interesting example is the parametric forced harmonic oscillator
with a time-dependent frequency $\omega (t),$ see \cite{Peremolov:Popov}
\begin{equation}
H(t)\psi =-\frac{1}{2}\triangle \psi +\frac{m}{2}\omega (t)^{2}x^{2}\psi
-f(t)x\psi .
\end{equation}%
This case is relevant in the context of charged particle traps \cite%
{Gheorghe}.

\subsection{Organization of the paper}

In Section 2 we derive the explicit formula for the fundamental solution of
\ (\ref{LStime}) with Hamiltonian (\ref{Hamiltonian}) (Lemma 1). We also
present explicitly the time evolution operator and relevant properties
(Corollary 1), conditions on the uniqueness of the solution for (\ref{LStime}%
)-(\ref{Hamiltonian}), and continuous dependence on the initial data and
smoothness of the solution (Theorem 1). In Section 3 we discuss Strichartz
type estimates for $U_{H}$ (Lemma 5 and Theorem 2). Finally, in Section 4 we
discuss local well-posedness in $L_{x}^{2}(\mathbb{R}^{d})$ for the
nonlinear case in the subcritical sense (Proposition 1).

\section{\textbf{LINEAR CASE}}

In this section we wish to prove a generalization of the formula (\ref%
{formula1})-(\ref{formula2}) in $\mathbb{R}^{d}$:

\begin{lemma}
(Fundamental Solution) 1. Let $\varphi \in S(\mathbb{R}^{d})$. The Cauchy
initial value problem (\ref{LStime})-(\ref{Hamiltonian}) has the following
fundamental solution%
\begin{equation}
G_{H}(x,y,t)=\left( \prod\limits_{j=2}^{d}\frac{1}{2\pi i\mu _{j}\left(
t\right) }\right) ^{d/2}e^{i\left( \sum \alpha _{j}\left( t\right)
x_{j}^{2}+\beta _{j}\left( t\right) x_{j}y_{j}+\gamma _{j}\left( t\right)
y_{j}^{2}+\delta _{j}\left( t\right) x_{j}+\varepsilon _{j}\left( t\right)
y_{j}+\kappa _{j}\left( t\right) \right) }.  \label{fundamental solution 1}
\end{equation}%
$\mu _{j}$ satisfies
\begin{equation}
\mu _{j}^{\prime \prime }+4\sigma _{j}\left( t\right) \mu _{j}=0,
\label{characteristic}
\end{equation}%
with $\sigma _{j}\left( t\right) =b_{j}(t)/2-c_{j}^{2}(t)/4-c_{j}^{\prime
}(t)/4,$ which must be solved subject to $\mu _{j}(0)=0$, $\mu _{j}^{\prime
}(0)=1$. Furthermore, $\alpha _{j}\left( t\right) ,$ $\beta _{j}\left(
t\right) ,$ $\gamma _{j}\left( t\right) ,$ $\delta _{j}\left( t\right) ,$ $%
\varepsilon _{j}\left( t\right) ,$ $\kappa _{j}\left( t\right) $ are
differentiable in time $t$ only and are given explicitly by
\begin{equation}
\alpha _{j}\left( t\right) =\frac{1}{2}\frac{\mu _{j}^{\prime }\left(
t\right) }{\mu _{j}\left( t\right) }-\frac{c_{j}(t)}{2},  \label{schr15}
\end{equation}%
\begin{equation}
\beta _{j}\left( t\right) =-\frac{1}{\mu _{j}\left( t\right) },
\label{scr16}
\end{equation}%
\begin{equation}
\gamma _{j}\left( t\right) =\frac{1}{2\mu _{j}\left( t\right) \mu
_{j}^{\prime }\left( t\right) }\ -2\int_{0}^{t}\frac{\sigma _{j}\left( \tau
\right) }{\left( \mu _{j}^{\prime }\left( \tau \right) \right) ^{2}}\ d\tau +%
\frac{c_{j}(0)}{2}  \label{schr17}
\end{equation}%
\begin{equation}
\delta _{j}\left( t\right) =\frac{1}{\mu _{j}\left( t\right) }%
\int_{0}^{t}\left( f_{j}\left( \tau \right) -c_{j}\left( \tau \right)
g_{j}\left( \tau \right) \right) \mu _{j}\left( \tau \right) +g_{j}\left(
\tau \right) \mu _{j}^{\prime }\left( \tau \right) \ d\tau ,  \label{schr18}
\end{equation}%
\begin{eqnarray}
\varepsilon _{j}\left( t\right) &=&-\frac{\delta _{j}\left( t\right) }{\mu
_{j}^{\prime }\left( t\right) }+4\int_{0}^{t}\frac{\mu _{j}\left( \tau
\right) \delta _{j}\left( \tau \right) \sigma _{j}\left( \tau \right) }{%
\left( \mu _{j}^{\prime }\left( \tau \right) \right) ^{2}}d\tau
\label{schr18a} \\
&&\quad +\int_{0}^{t}\frac{1}{\mu ^{\prime }\left( \tau \right) }\left(
f_{j}\left( \tau \right) -c_{j}\left( \tau \right) g_{j}\left( \tau \right)
\right) \ d\tau ,  \notag
\end{eqnarray}%
\begin{eqnarray}
\kappa _{j}\left( t\right) &=&\frac{\mu _{j}\left( t\right) }{2\mu
_{j}^{\prime }\left( t\right) }\delta _{j}^{2}\left( t\right) -2\int_{0}^{t}%
\frac{\sigma _{j}\left( \tau \right) }{\left( \mu _{j}^{\prime }\left( \tau
\right) \right) ^{2}}\left( \mu _{j}\left( \tau \right) \delta _{j}\left(
\tau \right) \right) ^{2}\ d\tau  \label{schr19} \\
&&\quad -\int_{0}^{t}\frac{\mu _{j}\left( \tau \right) \delta _{j}\left(
\tau \right) }{\mu _{j}^{\prime }\left( \tau \right) }\left( f_{j}\left(
\tau \right) -c_{j}\left( \tau \right) g_{j}\left( \tau \right) \right) \
d\tau  \notag
\end{eqnarray}%
with
\begin{equation}
\delta _{j}\left( 0\right) =g_{j}\left( 0\right) ,\qquad \varepsilon
_{j}\left( 0\right) =-\delta _{j}\left( 0\right) ,\qquad \kappa _{j}\left(
0\right) =0.  \label{conditions}
\end{equation}%
2. Convergence to the initial data:
\begin{equation*}
\lim_{t\rightarrow 0^{+}}\int_{\mathbb{R}^{d}}G_{H}(x,y,t)\psi (y,t)dy=\psi
(x,0).
\end{equation*}
\end{lemma}

Thus, the fundamental solution (propagator) is explicitly given by (\ref%
{fundamental solution}) in terms of the characteristic function (\ref%
{characteristic}) with (\ref{schr15})-(\ref{schr19}).

\begin{remark}
The conditions (\ref{conditions}), which are justified since we are looking
for the following asymptotic formula, see \cite{Sua:Sus2}, holds:
\begin{eqnarray*}
\frac{e^{i(\alpha _{j}\left( t\right) x_{j}^{2}+\beta _{j}\left( t\right)
x_{j}y_{j}+\gamma _{j}\left( t\right) y_{j}^{2}+\delta _{j}\left( t\right)
x_{j}+\varepsilon _{j}\left( t\right) y_{j}+\kappa _{j}\left( t\right) )}}{%
\sqrt{2\pi i\mu _{j}\left( t\right) }} &\rightarrow &\frac{1}{\sqrt{2\pi it}}%
\exp \left( i\frac{\left( x_{j}-y_{j}\right) ^{2}}{2t}\right) \exp \left(
ig_{j}\left( 0\right) \left( x_{j}-y_{j}\right) \right) \\
&&\times \exp \left( -\frac{ic_{j}(0)}{2}(x_{j}^{2}-y_{j}^{2})\right) ,\text{
}t\rightarrow 0^{+}.
\end{eqnarray*}
\end{remark}

\begin{theorem}
1. Let $\varphi \in S(\mathbb{R}^{d}),$ then the Cauchy initial value
problem (\ref{LStime})-(\ref{Hamiltonian}) has the following unitary
evolution operator:%
\begin{equation}
U_{H}(t)\varphi \equiv \left( \prod\limits_{j=2}^{d}\frac{1}{2\pi i\mu
_{j}\left( t\right) }\right) ^{d/2}\int_{\mathbb{R} ^{d}}e^{i\left( \sum
\alpha _{j}\left( t\right) x_{j}^{2}+\beta _{j}\left( t\right)
x_{j}y_{j}+\gamma _{j}\left( t\right) y_{j}^{2}+\delta _{j}\left( t\right)
x_{j}+\varepsilon _{j}\left( t\right) y_{j}+\kappa _{j}\left( t\right)
\right) }\varphi (y)dy.  \label{General Operator}
\end{equation}

2. If $\varphi \in S(\mathbb{R}^{d}),$ then $U_{H}(t)\varphi $ $\in S(%
\mathbb{R}^{d})$.

3. $U_{H}(t,s)=U_{H}(t)U_{H}^{-1}(s)$ and by duality $U_{H}(t,s)$ can be
extended to $S^{\prime }(\mathbb{R}^{d}).$ Furthermore, $U_{H}(\cdot
)\varphi \in C(\mathbb{R} ,S^{\prime }(\mathbb{R}^{d}))$ for every $\varphi
\in S^{\prime }(\mathbb{R}^{d}).$

If $\psi $ satisfies (\ref{LStime})-(\ref{Hamiltonian}) and it is smooth,
then:

4. The following estimates hold:%
\begin{equation}
\Vert U_{H}(t)\varphi \Vert _{L^{2}(\mathbb{R}^{d})}=\Vert \varphi \Vert
_{L^{2}(\mathbb{R}^{d})},  \label{Estimate1}
\end{equation}%
\begin{equation}
\left\Vert U_{H}\left( t,s\right) \varphi \right\Vert _{L^{\infty }(\mathbb{R%
}^d)}\leq \left( \prod\limits_{j=1}^{d}\frac{1}{\sqrt{4\pi i\mu _{j}\left(
t\right) \mu _{j}\left( s\right) \left( \gamma _{j}\left( s\right) -\gamma
_{j}\left( t\right) \right) }}\right) ^{d/2}\left\Vert \varphi \right\Vert
_{L^{1}(\mathbb{R}^{d})}.  \label{Estimate2}
\end{equation}%
5. Uniqueness and continuous dependence on the initial data in $L_{x}^{2}(%
\mathbb{R}^{d})$ holds.
\end{theorem}

\begin{remark}
The function $\mu _{j}$ will characterize the singularities \cite%
{Craig:Kappeler:Strauss}, \cite{Ka:Ro:Ya}, \cite{Zelditch} and \cite%
{Yajima1996}. More complicated cases may include special functions, like
Bessel, hypergeometric or elliptic functions. For a nice conection between
the characteristic equation (\ref{characteristic})\ and Ehrenfest Theorems
see \cite{Co:Suazo:Su2}.
\end{remark}

\begin{corollary}
The evolution operator associated to (\ref{LStime})-(\ref{Hamiltonian})
satisfies the following properties:

1. $U_{H}(t,s)=U_{H}(t)U_{H}^{-1}(s).$

2. $U_{H}(t,t)=Id.$

3. The map $(t,s)\rightarrow U_{H}(t,s)$ is strongly continuous.

4. $U_{H}(t,\tau )U_{H}(\tau ,s)=U_{H}(t,s).$
\end{corollary}

\subsection{Proof of the Lemma 1}

We follow \cite{Cor-Sot:Lop:Sua:Sus} where the fundamental solution is
constructed for a more general case in $d=1$ dimension. The Lemma follows
from the construction in the $1d$ case and then using tensor product to
construct the $d$-dimensional fundamental solution. We recall how to
construct the fundamental solution in the one dimensional case for%
\begin{equation}
\frac{\partial \psi _{j}}{\partial t}=-\frac{1}{2}\frac{\partial ^{2}\psi
_{j}}{\partial x_{j}^{2}}+\frac{b_{j}\left( t\right) }{2}x_{j}^{2}\psi
_{j}-f_{j}(t)x_{j}\psi _{j}+ig_{j}(t)\frac{\partial \psi _{j}}{\partial x_{j}%
}-i\frac{c_{j}\left( t\right) }{2}\left( 2x_{j}\frac{\partial \psi _{j}}{%
\partial x_{j}}-\psi _{j}\right) .
\end{equation}%
The fundamental solution is found by using the ansatz
\begin{equation}
\psi _{j}=A_{j}e^{iS_{j}}=A_{j}\left( t\right) e^{iS_{j}\left( x,y,t\right) }
\label{fundamental solution}
\end{equation}%
with
\begin{equation}
A_{j}=A_{j}\left( t\right) =\frac{1}{\sqrt{2\pi i\mu _{j}\left( t\right) }}
\label{schr3}
\end{equation}%
and
\begin{equation}
S_{j}=S_{j}\left( x,y,t\right) =\alpha _{j}\left( t\right) x_{j}^{2}+\beta
_{j}\left( t\right) x_{j}y_{j}+\gamma _{j}\left( t\right) y_{j}^{2}+\delta
_{j}\left( t\right) x_{j}+\varepsilon _{j}\left( t\right) y_{j}+\kappa
_{j}\left( t\right) ,  \label{schr4}
\end{equation}%
where $\alpha _{j}\left( t\right) ,$ $\beta _{j}\left( t\right) ,$ $\gamma
_{j}\left( t\right) ,$ $\delta _{j}\left( t\right) ,$ $\varepsilon
_{j}\left( t\right) ,$ and $\kappa _{j}\left( t\right) $ are differentiable
real-valued functions of time $t$ only. Indeed,
\begin{equation}
\frac{\partial S_{j}}{\partial t}=-\frac{1}{2}\left( \frac{\partial S_{j}}{%
\partial x}\right) ^{2}-b_{j}x_{j}^{2}+f_{j}x_{j}+\left(
g_{j}-c_{j}x_{j}\right) \frac{\partial S_{j}}{\partial x_{j}}  \label{schr5}
\end{equation}%
by choosing
\begin{equation}
\frac{\mu _{j}^{\prime }}{2\mu _{j}}=\frac{1}{2}\frac{\partial ^{2}S_{j}}{%
\partial x_{j}^{2}}+\frac{c_{j}}{2}=\alpha _{j}\left( t\right) +\frac{%
c_{j}(t)}{2}.  \label{schr6}
\end{equation}%
Equating the coefficients of all admissible powers of $x_{j}^{m}y_{j}^{n}$
with $0\leq m+n\leq 2$ gives the following system of ordinary differential
equations:

\begin{align}
& \frac{d\alpha _{j}}{dt}+b_{j}\left( t\right) +2c_{j}\left( t\right) \alpha
_{j}+2\alpha _{j}^{2}=0,  \label{schr7} \\
& \frac{d\beta _{j}}{dt}+\left( c_{j}\left( t\right) +2\alpha _{j}\left(
t\right) \right) \beta _{j}=0,  \label{schr8} \\
& \frac{d\gamma _{j}}{dt}+\frac{\beta _{j}^{2}\left( t\right) }{2}=0,
\label{schr9} \\
& \frac{d\delta _{j}}{dt}+\left( c_{j}\left( t\right) +2\alpha _{j}\left(
t\right) \right) \delta _{j}=f_{j}\left( t\right) +2\alpha _{j}\left(
t\right) g_{j}\left( t\right) ,  \label{schr10} \\
& \frac{d\varepsilon _{j}}{dt}=\left( g_{j}\left( t\right) -\delta
_{j}\left( t\right) \right) \beta _{j}\left( t\right) ,  \label{schr11} \\
& \frac{d\kappa _{j}}{dt}=g_{j}\left( t\right) \delta _{j}\left( t\right) -%
\frac{\delta _{j}^{2}\left( t\right) }{2},  \label{schr12}
\end{align}%
where the first equation is the Riccati nonlinear differential equation.
Substituting (\ref{schr6}) into (\ref{schr7}) results in the second order
linear equation
\begin{equation}
\mu _{j}^{\prime \prime }+4\sigma _{j}\left( t\right) \mu _{j}=0
\label{schr13}
\end{equation}%
with
\begin{equation}
\sigma _{j}\left( t\right) =\frac{b_{j}(t)}{2}-\frac{c_{j}^{2}(t)}{4}-\frac{%
c_{j}^{\prime }}{4},  \label{schr13a}
\end{equation}%
which must be solved subject to the initial data
\begin{equation}
\mu _{j}\left( 0\right) =0,\qquad \mu _{j}^{\prime }\left( 0\right) =1.
\label{schr13b}
\end{equation}%
We shall refer to equation (\ref{schr13}) as the \textit{characteristic
equation} and its solution $\mu _{j}\left( t\right) ,$ subject to (\ref%
{schr13b}), as the \textit{characteristic function.} \ Using integration by
parts we can solve (\ref{schr7})-(\ref{schr12}) obtaining (\ref{schr15})-(%
\ref{schr19}).

Part 2 is a consequence of the two following results in the one dimensional
case proven in \cite{Sua:Sus2}.

\begin{lemma}
Let $G$ be defined by (\ref{fundamental solution 1}). There exists a
complex-valued function $K$ satisfying%
\begin{equation}
\int_{\mathbb{R}^{d}}\int_{\mathbb{R}^{d}}G(x,z,t)K(z,y,0)\chi (y)dzdy=\int_{%
\mathbb{R}^{d}}\int_{\mathbb{R}^{d}}G(x,z,t)K(z,y,0)\chi (y)dydz
\end{equation}%
and
\begin{equation}
\int_{\mathbb{R}^{d}}G(x,y,t)\psi (y,t)dy=\int_{\mathbb{R}^{d}}K(x,y,t)\chi
(y)dy{\tiny .}
\end{equation}
\end{lemma}

\begin{lemma}
If we consider the intial data $\psi (x,0)$ such that $\psi (x,0)=\int_{%
\mathbb{R}^{d}}K(x,y,0)\chi (y)dy$ for some $\chi \in L^{1}(\mathbb{R}^{d}),$
then%
\begin{equation*}
\lim_{t\rightarrow 0^{+}}\int_{\mathbb{R}^{d}}G(x,y,t)\psi (y,t)dy=\psi
(x,0).
\end{equation*}
\end{lemma}

\begin{remark}
For the case of the free particle propagator $K(x,y,0)=e^{ix\cdot y}$ and $%
\chi (y)$ is the Fourier transform of $\psi (x,0).$
\end{remark}

\subsection{Proof of Theorem}

1. The propagator $U_{H}(t)$ for the equation (\ref{LStime})-(\ref%
{Hamiltonian}) can be written as
\begin{equation}
U_{H}(t)\varphi =A_{t}B_{t}\mathfrak{F}\left( C_{t}\varphi \right) ,
\end{equation}%
where $A_{t}(x)=e^{i(\alpha \left( t\right) x^{2}+\delta \left( t\right)
x+\kappa \left( t\right) )},$ $C_{t}(x)=e^{i(\gamma \left( t\right)
x^{2}+\varepsilon \left( t\right) x)},$ $B_{t}w(x)=\left( 2\pi i\mu
(t)\right) ^{-\frac{1}{2}}w\left( -\beta (t)x/2\pi \right) $ and $\mathfrak{F%
}$ is the Fourier transform. Since the Fourier transform is an isomorphism
on the Schwartz space we have the operator $U_{H}(t)$ is an isomorphism on
the Schwartz space.

2. We claim that
\begin{equation}
\psi \left( x,0\right) =U_{H}^{-1}\left( t\right) \psi \left( x,t\right)
=\int_{-\infty }^{\infty }H\left( x,y,t\right) \ \psi \left( y,t\right) \ dy,
\end{equation}%
where
\begin{equation}
H\left( x,y,t\right) =\left( \prod\limits_{j=1}^{d}\frac{1}{-2\pi i\mu
_{j}\left( t\right) }\right) ^{d/2}e^{-i\sum_{j=1}^{d}S_{j}\left(
y_{j},x_{j},t\right) }
\end{equation}%
such that%
\begin{equation}
U_{H}\left( t\right) U_{H}^{-1}\left( t\right) =U_{H}^{-1}\left( t\right)
U_{H}\left( t\right) =I=\text{id}.  \label{inv3}
\end{equation}%
First we observe that the following orthogonality relations of the kernels
hold:%
\begin{eqnarray}
\int_{\mathbb{R}^{d}}G\left( x,y,t\right) H\left( y,z,t\right) \ dy
&=&e^{i\sum_{j=1}^{d}\left( \alpha _{j}\left( t\right) \left(
x_{j}+z_{j}\right) +\delta _{j}\left( t\right) \right) \left(
x_{j}-z_{j}\right) }\prod\limits_{j=1}^{d}\delta \left( x_{j}-z_{j}\right) ,
\label{inv6} \\
\int_{\mathbb{R}^{d}}H\left( x,y,t\right) G\left( y,z,t\right) \ dy
&=&e^{-i\sum_{j=1}^{d}\left( \gamma _{j}\left( t\right) \left(
x_{j}+z_{j}\right) +\varepsilon _{j}\left( t\right) \right) \left(
x_{j}-z_{j}\right) }\prod\limits_{j=1}^{d}\delta \left( x_{j}-z_{j}\right) ,
\label{inv7}
\end{eqnarray}%
where $\delta \left( x\right) $ is the Dirac delta function with respect to
the space coordinates.

Next, we have%
\begin{eqnarray*}
&&U_{H}^{-1}\left( t\right) U_{H}\left( t\right) \psi \left( x,0\right)
=U_{H}^{-1}\left( t\right) \psi \left( x,t\right) \\
&&\quad =\int_{\mathbb{R}^{d}}H\left( x,y,t\right) \ \psi \left( y,t\right)
\ dy \\
&&\quad =\int_{\mathbb{R}^{d}}H\left( x,y,t\right) \ \left( \int_{\mathbb{R}%
^{d}}G_{H}\left( y,z,t\right) \ \psi \left( z,0\right) \ dz\right) \ dy \\
&&\quad =\int_{\mathbb{R}^{d}}\left( \int_{\mathbb{R}^{d}}H\left(
x,y,t\right) G_{H}\left( y,z,t\right) \ dy\right) \ \psi \left( z,0\right) \
dz \\
&&\quad =\int_{\mathbb{R}^{d}}e^{-i\sum_{j=1}^{d}\left( \gamma _{j}\left(
t\right) \left( x_{j}+z_{j}\right) +\varepsilon _{j}\left( t\right) \right)
\left( x_{j}-z_{j}\right) }\prod\limits_{j=1}^{d}\delta \left(
x_{j}-z_{j}\right) \ \psi \left( z,0\right) \ dz \\
&&\quad =\psi \left( x,0\right) ,
\end{eqnarray*}%
or $U_{H}^{-1}\left( t\right) U_{H}\left( t\right) =I.$ A formal proof of
the second relation $U_{H}\left( t\right) U_{H}^{-1}\left( t\right) =I$ is
similar. The rest of the statement follows for a standard duality argument,
see \cite{Caz}.

3. This claim is a consequence of $(1)$ multiplying the equation%
\begin{equation*}
i\psi _{t}=-a(t)\psi _{xx}+b(t)x^{2}\psi -f(t)x\psi -i(c(t)x-g(t))\partial
_{x}\psi -i\frac{c\left( t\right) }{2}\psi
\end{equation*}%
by $\overline{\psi }$ $(2)$ integrating in the space variable$,$ and $(3)$
equating the imaginary parts of both sides to obtain%
\begin{equation}
Re\int \psi _{t}\overline{\psi }dx=-\frac{1}{2}\int \left( c(t)x-g(t)\right)
\partial _{x}(|\psi |^{2})dx-\frac{1}{2}\int c\left( t\right) |\psi |^{2}dx;
\end{equation}%
the estimate now follows from the solution being smooth.

We introduce the integral operator $U_{H}\left( t,s\right) =U_{H}\left(
t\right) U_{H}^{-1}\left( s\right) $ by%
\begin{equation}
U_{H}\left( t\right) U_{H}^{-1}\left( s\right) \psi \left( x,s\right)
=\int_{-\infty }^{\infty }G\left( x,y,t,s\right) \psi \left( y,s\right) \ dy
\label{inv8a}
\end{equation}%
with the kernel given by%
\begin{equation}
G\left( x,y,t,s\right) =\int_{\mathbb{R}^{d}}G_{H}\left( x,z,t\right)
H\left( z,y,s\right) \ dz.  \label{inv8}
\end{equation}%
Here,%
\begin{eqnarray}
&&G\left( x,y,t,s\right) =\left( \prod\limits_{j=1}^{d}\frac{1}{\sqrt{4\pi
i\mu _{j}\left( t\right) \mu _{j}\left( s\right) \left( \gamma _{j}\left(
s\right) -\gamma _{j}\left( t\right) \right) }}\right) ^{d/2}  \label{inv9a}
\\
&&\qquad \times \sum_{j=1}^{d}\exp \left( i\left( \alpha _{j}\left( t\right)
x_{j}^{2}-\alpha _{j}\left( s\right) y_{j}^{2}+\delta _{j}\left( t\right)
x_{j}-\delta _{j}\left( s\right) y_{j}+\kappa _{j}\left( t\right) -\kappa
_{j}\left( s\right) \right) \right)  \notag \\
&&\qquad \times \sum_{j=1}^{d}\exp \left( \frac{\left( \beta _{j}\left(
t\right) x_{j}-\beta _{j}\left( s\right) y_{j}+\varepsilon _{j}\left(
t\right) -\varepsilon _{j}\left( s\right) \right) ^{2}}{4i\left( \gamma
_{j}\left( t\right) -\gamma _{j}\left( s\right) \right) }\right) .  \notag
\end{eqnarray}

\begin{eqnarray*}
\left\vert U_{H}\left( t,s\right) \psi \left( x,s\right) \right\vert
&=&\left\vert \int_{\mathbb{R}^{d}}G\left( x,y,t,s\right) \psi \left(
y,s\right) \ dy\right\vert \\
&\leq &\left( \prod\limits_{j=1}^{d}\frac{1}{\sqrt{4\pi i\mu _{j}\left(
t\right) \mu _{j}\left( s\right) \left( \gamma _{j}\left( s\right) -\gamma
_{j}\left( t\right) \right) }}\right) ^{d/2}\int_{\mathbb{R}^{d}}\left\vert
\psi \left( y,s\right) \right\vert \ dy.
\end{eqnarray*}%
Thus, the estimate (\ref{Estimate2}) holds.

4. The uniqueness of the solution and continuous dependence on the initial
data follows by standard arguments using estimates of the type of (\ref%
{Estimate1}).

\section{\textbf{EXAMPLES OF STRICHARTZ TYPE ESTIMATES}}

In this section we will discuss Strichartz type estimates for the operator $%
U_{H}.$

\begin{definition}
We say that the exponent pair $(q,r)$ is $\sigma -admissible$ if $q,r\geq 2$%
, $(q,r,\sigma )$ $\neq $ $(2,\infty ,1)$ and
\begin{equation*}
\frac{1}{q}+\frac{\sigma }{r}\leq \frac{\sigma }{2}.
\end{equation*}%
If equality holds we say that $(q,r)$ is sharp $\sigma $-admissible,
otherwise, we say that $(q,r)$ is nonsharp $\sigma $-admissible. Note, in
particular, that when $\sigma >1$ the endpoint
\begin{equation*}
P=\left( 2,\frac{2\sigma }{\sigma -1}\right)
\end{equation*}%
is sharp $\sigma $-admissible.
\end{definition}

The following inequalities are known as Strichartz estimates studied by
Strichartz, Ginibre, Velo, Keel and Tao and others, see \cite{Keel} for the
following version.

\begin{lemma}
If $U(t):H\rightarrow L^{2}(x)$, where $H$ \ is a Hilbert space and

\begin{itemize}
\item For all $t>0$ and $f\in H,$ we have
\begin{equation}
||U(t)f||_{L_{x}^{2}}\lesssim ||f||_{H},
\end{equation}

\item For some $\sigma >0,\ t\neq s$ and $g\in L^{1}(x)$,%
\begin{equation}
||U(t)U(s)^{\ast }g||_{\infty }\lesssim |t-s|^{-\sigma }||g||_{L^{1}}.
\label{untruncated decay}
\end{equation}%
Then
\begin{eqnarray}
||U(t)f||_{L_{t}^{q}L_{x}^{r}} &\lesssim &||f||_{H}  \label{(5)} \\
\left\vert \left\vert \int \left( U(s)\right) ^{\ast }F(s)ds\right\vert
\right\vert _{L^{2}} &\lesssim &||F||_{L_{t}^{q^{\prime }}L_{x}^{r^{\prime
}}}  \label{(6)} \\
\left\vert \left\vert \int_{s<t}U(t)\left( U(s)\right) ^{\ast
}F(s)ds\right\vert \right\vert _{_{L_{t}^{q}L_{x}^{r}}} &\lesssim
&||F||_{L_{t}^{\tilde{q}^{\prime }}L_{x}^{\tilde{r}^{\prime }}}  \label{(7)}
\end{eqnarray}%
hold for all sharp $\sigma $-admissible exponent pairs $(q,r)$, $(\tilde{q},%
\bar{r}).$
\end{itemize}
\end{lemma}

In order to obtain inequalities (\ref{(5)}), (\ref{(6)}), (\ref{(7)}) for
our operator $U_{H},$ we observe that the inequalities are valid if we
replace (\ref{untruncated decay}) by
\begin{equation}
||U(t)\left( U(s)\right) ^{\ast }F(s)||_{L_{x}^{\infty }}\leq w(t-s)^{\gamma
}||F(s)||_{L_{x}^{1}},\qquad w\in L_{\omega }^{1}.
\label{inequality with L1 weak function}
\end{equation}%
We can just mimic the proof of Lemma 3, see \cite{Tao} and the adaptation in
\cite{Carles} done for the case $U_{V}$. For our case we deal with the
inequality (\ref{inequality with L1 weak function}) and avoid the use of
semigroup properties that we do not have. More specifically we obtain:

\begin{lemma}
If $U_{H}(t)$ satisfies (\ref{inequality with L1 weak function}), then for
any $T\in \bar{\mathbb{R}}_{+}$,
\begin{eqnarray}
||U_{H}(t)f||_{L_{t}^{q}L_{x}^{r}} &\leq &c_{q}||w\mathbf{\cdot 1}%
_{(-2T,2T)}||_{L^{2}}^{\gamma _{1}}||f||_{L^{2}},  \label{(5)''} \\
\left\vert \left\vert \int \left( U_{H}(s)\right) ^{\ast }F(s)ds\right\vert
\right\vert _{L^{2}} &\lesssim &\tilde{c}_{\tilde{q}}||w\mathbf{\cdot 1}%
_{(-2T,2T)}||_{L^{2}}^{\gamma _{2}}||F||_{L_{t}^{q^{\prime
}}L_{x}^{r^{\prime }}},  \label{(6)"} \\
\left\vert \left\vert \int_{s<t}U_{H}(t)\left( U_{H}(s)\right) ^{\ast
}F(s)ds\right\vert \right\vert _{L_{t}^{q}((-T,T);L_{x}^{r})} &\leq &C_{q,%
\tilde{q}}||w\mathbf{\cdot 1}_{(-2T,2T)}||_{L^{2}}^{\gamma
_{3}}||F||_{L_{t}^{\tilde{q}^{\prime }}L_{x}^{\tilde{r}^{\prime }}},
\label{(7)"}
\end{eqnarray}%
holding for all sharp $\sigma $-admisible exponent pairs $(q,r)$, $(\tilde{q}%
,\bar{r}),$ where $\gamma _{1},$ $\gamma _{2}$ $,\gamma _{3}$ are constants$%
. $
\end{lemma}

For the sake of clarity we outline the proof:

First, observe that (\ref{(6)})\ implies (\ref{(5)})\ by duality. Therefore
we prove (\ref{(6)})
\begin{equation*}
\left\vert \left\vert \int U_{H}(s)^{\ast }F(s)ds\right\vert \right\vert
_{L^{2}(\mathbb{R}^{2})}\lesssim ||F||_{L^{q^{\prime }}(\mathbb{R},\
L^{r^{\prime }}(\mathbb{R}^{2}))}.
\end{equation*}%
By the TT* method this can be implied by the following inequality:
\begin{equation*}
\left\vert \int \int <U_{H}(s)^{\ast }F(s),U_{H}(t)^{\ast
}G(t)dsdt>\right\vert \lesssim ||F||_{L_{t}^{q^{\prime }}L_{x}^{r^{\prime
}}}||G||_{L_{t}^{q^{\prime }}L_{x}^{r^{\prime }}}.
\end{equation*}%
By symmetry, it suffices to prove%
\begin{equation}
|T(F,G)|\leq ||F||_{L_{t}^{q^{\prime }}L_{x}^{r^{\prime
}}}||G||_{L_{t}^{q^{\prime }}L_{x}^{r^{\prime }}}
\end{equation}%
where
\begin{equation*}
|T(F,G)|=\int \int_{s<t}<U_{H}(s)^{\ast }F(s),U_{H}(t)^{\ast }G(t)>dsdt.
\end{equation*}%
Since $U_{H}$ is unitary ($U_{H}^{\ast }=U_{H}^{-1}$) on $L^{2},$ by
Holder's inequality and the energy estimate, we get%
\begin{eqnarray}
\left\vert <U_{H}(s)^{\ast }F(s),U_{H}(t)^{\ast }G(t)>\right\vert &\leq
&||U_{H}(s)^{\ast }F(s)||_{L_{x}^{2}}||U_{H}(t)^{\ast }G(t)||_{L_{x}^{2}} \\
&=&||F(s)||_{L_{x}^{2}}||G(s)||_{L_{x}^{2}}.
\end{eqnarray}%
By assumption (\ref{inequality with L1 weak function}) and from the above
estimate, we get%
\begin{eqnarray}
\left\vert <U_{H}(s)^{\ast }F(s),U_{H}(t)^{\ast }G(t)>\right\vert &\leq
&||U_{H}(t)U_{H}(s)^{\ast }F(s)||_{L_{x}^{\infty }}||G(t)||_{L_{x}^{1}} \\
&\lesssim &\omega (s-t)^{\gamma }||F(s)||_{L_{x}^{1}}||G(s)||_{L_{x}^{1}}.
\end{eqnarray}%
The last equality follows by Holder's inequality and $\gamma $ is coming
from (\ref{inequality with L1 weak function}).

Now by interpolating with%
\begin{equation*}
\left\vert <U_{H}(s)^{\ast }F(s),U_{H}(t)^{\ast }G(t)>\right\vert \leq
||F(s)||_{L_{x}^{2}}||G(s)||_{L_{x}^{2}},
\end{equation*}%
we obtain that if $r^{\prime }$ is defined by $1/r^{\prime }=1-\theta
+\theta /2,\ 0<\theta <1$, then $1<r^{\prime }<2.$ If we denote by $r$ the
dual exponent of $r^{\prime }$ then $1/r=\theta /2,$ and choosing $q$ such
that $(q,r)$ is $\gamma $-admissible, we will also get that $(q^{\prime
},r^{\prime })$ is sharp $\gamma $-admissible, and that $\gamma (1-\theta
)=2/q.$ Therefore,%
\begin{equation*}
\left\vert <U_{H}(s)^{\ast }F(s),U_{H}(t)^{\ast }G(t)>\right\vert \leq
\omega (t-s)^{\frac{2}{q}}||F(s)||_{L_{x}^{r^{\prime
}}}||G(s)||_{L_{x}^{r^{\prime }}}.
\end{equation*}%
Since $(q,r)$ is sharp $\gamma $-admissible (and it is not an endpoint),
then we can apply weak young inequality, obtaining%
\begin{eqnarray*}
\int \int \left\vert <U_{H}(s)^{\ast }F(s),U_{H}(t)^{\ast }G(t)>\right\vert
dsdt &\leq &\int \int \omega (t-s)^{\frac{2}{q}}||F(s)||_{L_{x}^{r^{\prime
}}}||F(s)||_{G_{x}^{r^{\prime }}}dsdt \\
&\leq &||\omega (t-s)||_{L_{\omega }^{\frac{q}{2}}}||F||_{L_{t}^{q^{\prime
}}L_{x}^{r^{\prime }}}||G||_{L_{t}^{q^{\prime }}L_{x}^{r^{\prime }}},
\end{eqnarray*}%
where $2/q+1/q^{\prime }+1/q^{\prime }=2,$ and since $\omega $ is in weak $%
L^{1}$, (\ref{(6)}) follows.

Now to prove (\ref{(7)}) we can proceed as in \cite{Tao} (Section 7).

The following result gives us conditions on operator $U_{H}$ to obtain
global in time Strichartz estimates:

\begin{theorem}
1. Consider the following restriction on the coefficients%
\begin{equation}
\frac{b_{j}(t)}{2}-\frac{c_{j}^{2}(t)}{4}-\frac{c_{j}^{\prime }(t)}{4}%
=\sigma _{j,\qquad }j\geqslant 1,\qquad \sigma _{j}\in \{-\frac{1}{4},0,%
\frac{1}{4}\}.  \label{condition on the coefficients1}
\end{equation}%
The evolution operator $U_{H}$ associated to the Cauchy problem (\ref{LStime}%
)-(\ref{Hamiltonian}) is given by (\ref{General Operator}) where%
\begin{equation}
\mu _{j}(t)=\left\{
\begin{array}{c}
\frac{\sinh (\omega _{j}t)}{\omega _{j}}\text{ , if }\sigma _{j}=-1, \\
t\text{ , \ \ \ \ \ \ if }\sigma _{j}=0, \\
\frac{\sin (\omega _{j}t)}{\omega _{j}}\text{ , if }\sigma _{j}=+1,%
\end{array}%
\right.  \label{Condition de mu}
\end{equation}%
and it satisfies
\begin{equation}
\left\Vert U_{H}\left( t,s\right) \varphi \right\Vert _{\infty }\leq \left(
\prod\limits_{j=1}^{d}\frac{1}{2\pi |\mu _{j}\left( t-s\right) |}\right)
^{d/2}\left\Vert \varphi \right\Vert _{1}.  \label{Estimate3}
\end{equation}%
Furthermore, if $\delta _{j}=-1$ for some $j,$ we have global in time
Strichartz estimates.
\end{theorem}

\begin{proof}
It is easy to see that (\ref{Estimate3}) follows from (\ref{Estimate2}) and (%
\ref{Condition de mu}). The global in time Strichartz estimates follow from
Lemma 5 and observing that (as was pointed out in \cite{Carles} (Section 2))
if $\delta _{k}=-1$ for some $k$ and $\delta _{j}=1$ for $i\neq k$ (the
worst of the possible cases), then
\begin{equation}
w(t)=C\left( \frac{1}{|t|}1_{|t|\leq \delta }+\left( e^{-\omega
_{k}t}\prod\limits_{j\neq k}^{d}\frac{1}{2\pi |\sin (\omega _{j}t)|}\right)
^{\frac{1}{d}}1_{|t|>\delta }\right)
\end{equation}%
is in $L_{\omega }^{1}(\mathbb{R}),$ and from Lemma 5 we obtain global in
time Strichartz estimates$.$
\end{proof}

\section{\textbf{DAMPED HARMONIC NONLINEAR SCHR\"{O}DINGER EQUATION}}

All linear exactly solvable models which we discussed in Sections 1 and 2
are of interest in a general treatment of the nonlinear time-dependent Schr%
\"{o}dinger equation, see \cite{Howland}, \cite{Jafaev}, \cite{Naibo:Stef},
\cite{Rod:Schlag}. In this Section we apply the results of the last sections
to the study of the nonlinear version of equation (\ref{LStime})-(\ref%
{Hamiltonian}) (with algebraic nonlinearity):%
\begin{equation}
i\frac{\partial u}{\partial t}=-\frac{1}{2}\Delta u+\sum_{j=1}^{d}\left(
\frac{b_{j}\left( t\right) }{2}x_{j}^{2}u-f_{j}(t)x_{j}u+ig_{j}(t)\frac{%
\partial u}{\partial x_{j}}-i\frac{c_{j}\left( t\right) }{2}\left( 2x_{j}%
\frac{\partial u}{\partial x_{j}}-u\right) \right) +h|u|^{p-1}u
\label{nonlinearcase1}
\end{equation}%
\begin{equation}
u(x,0)=u_{0}(x).  \label{nonlinearcaseinitial1}
\end{equation}

It includes the damped harmonic nonlinear Schr\"{o}dinger equation (\ref%
{damped harmonic nonlinear Schr}) and the following well-known cases:

\begin{itemize}
\item The nonlinear Schr\"{o}dinger equation with zero potential%
\begin{equation}
i\frac{\partial u}{\partial t}=-\frac{1}{2}\bigtriangleup u+hu|u|^{p-1}.
\end{equation}

\item The nonlinear Schr\"{o}dinger equation with the quadratic potential
possibly depending on time \cite{Carles}, \cite{Carlesstark}%
\begin{equation}
i\frac{\partial u}{\partial t}=-\frac{1}{2}\bigtriangleup u+\sum_{j=1}^{d}%
\frac{b_{j}\left( t\right) }{2}x_{j}^{2}u+hu|u|^{p-1}.
\end{equation}

\item The Gross--Pitaevskii equation%
\begin{equation}
i\frac{\partial u}{\partial t}=-\frac{1}{2}\triangle u+\frac{m}{2}\omega
(t)^{2}x^{2}u-f(t)xu+hu|u|^{p-1},
\end{equation}%
see \cite{Pitaevskii}, \cite{Yuri}, \cite{Perez}.
\end{itemize}

These equations have different applications, for example, the propagation of
waves, optical transmission lines with online modulators, propagation of
light beams in nonlinear media with a gradient of the refraction index, or
in the theory of Bose-Einstein condensate in trapped gases \cite{Pitaevskii}%
, \cite{Yuri}. The following local well-posedness result in $L_{x}^{2}(%
\mathbb{R}^{d})$ in the subcritical sense is a consequence of Strichartz
estimates and a boostrap argument \cite{Tao}:

\begin{proposition}
Let $p$ be an $L_{x}^{2}-$subcritical exponent ($0<p-1<4/d$), $h=\pm 1.$
Then for any $R>0$ there exists $T>0$ such that for all $u_{0}\in L_{x}^{2}(%
\mathbb{R}^{d})$ in the ball $B_{R}=\{u_{0}\in L_{x}^{2}(\mathbb{R}%
^{d}):\left\Vert u_{0}\right\Vert _{L_{x}^{2}(\mathbb{R}^{d})}<R\}$ there
exists a unique strong $L_{x}^{2}$ solution $u$\ to\ (\ref{nonlinearcase1})-(%
\ref{nonlinearcaseinitial1}) in the space $S^{0}([-T,T]\times \mathbb{R}%
^{d})\subset $\ $C_{t}^{0}L_{x}^{2}([-T,T]\times \mathbb{R}^{d}).$
\end{proposition}

\textit{Sketch of the proof: }Since we obtained Strichartz estimates, see
Section 3, we can follow the proof for the case of the nonlinear Schr\"{o}%
dinger equation without potential, see \cite{Tao}, \cite{Caz} or \cite{Sulem}%
.

\begin{remark}
(Time dependent nonlinearity) One can also consider a generalization of the
nonautonomous Schr\"{o}dinger equation
\begin{equation}
i\frac{\partial u}{\partial t}=-\frac{1}{2}\Delta u+\sum_{j=1}^{d}\left(
\frac{b_{j}\left( t\right) }{2}x_{j}^{2}u-f_{j}(t)x_{j}u+ig_{j}(t)\frac{%
\partial u}{\partial x_{j}}-i\frac{c_{j}\left( t\right) }{2}\left( 2x_{j}%
\frac{\partial u}{\partial x_{j}}-u\right) \right) +h(t)|u|^{p-1}u
\end{equation}%
\begin{equation}
u(x,0)=u_{0}(x),
\end{equation}%
considering for example $h\in C^{\infty }(\mathbb{R};\mathbb{R}).$This
equation will include the nonautonomous Schr\"{o}dinger equation \cite{Caz}%
\begin{equation}
i\frac{\partial u}{\partial t}=-\frac{1}{2}\triangle u+h(t)u|u|^{p-1}.
\end{equation}
\end{remark}

\textbf{Acknowledgments.} The author is grateful to Dr Sergei K. Suslov for
his fundamental guidance and support and Dr. Svetlana Roudenko for very
useful discussions and key references. The author thanks Dr Luca Fanelli and
Dr Don Jones for helpful discussions and Dr. Carlos Castillo-Ch\'{a}vez for
his endeavours to open new research opportunities at ASU. Finally, the
author is indebted to Professor George A. Hagedorn, for pointing out
references relevant to this paper and future research.

\end{document}